\documentclass{sigplanconf}

\usepackage[T1]{fontenc}
\usepackage{ae,aecompl}
\usepackage[english]{babel}
\usepackage[psamsfonts]{amsfonts}
\usepackage{stmaryrd,amsmath,amssymb,amsbsy}
\usepackage{euscript}
\usepackage{graphicx}
\usepackage{color}
\usepackage{index}
\usepackage{url}
\usepackage{multirow}

\newbox\blindbox \newdimen\blindlen

\def\astree{{\sc Astr\'ee}}

\def\mc#1{{\mathcal{#1}}}

\def\mB#1{{\mathbb{#1}}}

\def\mr#1{{\mathrm{#1}}}
\def\mi#1{{\text{\it #1}}}
\def\mt#1{{\text{\tt #1}}}

\def\Id{\mi{Id}}
\def\N{\mB{N}}
\def\Z{\mB{Z}}
\def\R{\mB{R}}
\def\D{\mc{D}}
\def\V{\mc{V}}
\def\Val{\mB{V}}
\def\P{\mc{P}}
\def\C{\mc{C}}
\def\B{\mc{B}}
\def\F{\mc{F}}
\def\nullptr{\text{\o}}
\def\err{\omega}

\def\s{\sharp}

\def\deq{\;\stackrel{\mbox{{\rm\tiny def}}}{=}\;}

\def\widen{\mathbin{\triangledown}}

\def\lb#1{\llbracket\,#1\,\rrbracket}
\def\lc#1{{\{\hskip-0.25em|}\,#1\,{|\hskip-0.25em\}}}

\def\hi{\discretionary{}{}{}}

\def\vdots{\vbox{\baselineskip4pt\hbox{.}\hbox{.}\hbox{.}}}

\begin{document}


\conferenceinfo{LCTES'06} {June 14--16, 2006, Ottawa, Ontario, Canada.}
\copyrightyear{2006}
\copyrightdata{1-59593-362-X/06/0006}


\title{Field-Sensitive Value Analysis of Embedded C Programs with Union Types 
and Pointer Arithmetics}

\authorinfo{Antoine Min\'e}
           {\'Ecole Normale Sup\'erieure, Paris, France}
           {mine@di.ens.fr}

\maketitle


\begin{abstract}
We propose a memory abstraction able to lift existing numerical static analyses
to C programs containing union types, pointer casts, and arbitrary pointer 
arithmetics.
Our framework is that of a combined points-to and data-value analysis. 
We abstract the contents of compound variables in a field-sensitive way,
whether these fields contain numeric or pointer values, and use stock 
numerical abstract domains to find an overapproximation of 
all possible memory states---with the ability to discover relationships between
variables.
A main novelty of our approach is the dynamic mapping scheme we use to associate
a flat collection of abstract cells of scalar type
to the set of accessed memory 
locations, while taking care of byte-level aliases---{\em i.e.,} C variables
with incompatible types allocated in overlapping memory locations.
We do not rely on static type information which can be misleading in C
programs as it does not account for all the uses a memory zone may be put to.

Our work was incorporated within the \astree\ static analyzer that checks for
the absence of run-time-errors in embedded, safety-critical, numerical-intensive
software.
It replaces the former memory domain limited to well-typed, union-free,
pointer-cast free data-structures.
Early results demonstrate that this abstraction allows analyzing a larger
class of C programs, without much cost overhead.
\end{abstract}





\section{Introduction}

In embedded critical software, the slightest programming error can have the
most disastrous consequences.
Even when the high-level specification of a software is correct, its
actual implementation using efficient but unsafe low-level languages can 
introduce new kinds of bugs, such as run-time errors triggered by 
integer wrap-around or invalid floating-point operations---witness the demise 
of the Ariane launcher in 1996 \cite{ariane}.
Hence, there is a demand for tools able to check for potential run-time
errors in low-level programs, in an automatic and sound way.
To this end, we focus here on deriving the set of values the variables of a C
program can take during all its executions. 
We allow sound but generally 
incomplete approximations to ensure an efficient analysis.
This way, we are able to report a set of alarms that encompasses 
{\em all\/} possible
run-time error situations.
Hopefully, when the analysis is sufficiently precise, there are zero alarms
which actually {\em proves formally\/} the absence of run-time errors.

Unfortunately, the weak type system of the C programming language 
complicates value analysis greatly. 
In the presence of union types, pointer arithmetics
or pointer casts, the same sequence of memory bytes can be manipulated as 
values of distinct types.
Most existing analyses avoid the problem by either restricting the
input language or by being overly conservative about the contents of
memory locations that can be accessed with incompatible types---{\it i.e.},
treat them in a field-insensitive way.
We found these solutions to be insufficient to analyze actual embedded C 
codes provided by industrial end-users. 
As they exploit some knowledge of the bit-representation
of values and the low-level semantics of operators, tracking
precisely the manipulated values is required to prove the absence of
run-time errors.

To address these problems, we propose a field-sensitive value analysis for C
programs containing union types, pointer arithmetics and pointer casts.
Our main contribution is an abstraction that maps the memory, viewed as
untyped spans of bytes, to a collection of synthetic cells with integer or 
floating-point type.
We then rely on existing alias-unaware numerical analyses---such as intervals
\cite{ai} or octagons \cite{mine:oct}---to infer numerical invariants
on cells.
Our abstraction translates operations on byte-based memory locations 
into operations on cells, taking care of byte-level aliasing between cells.
We use a dynamic mapping because, due to pointer casts, the uses of the 
memory cannot be deduced from the static type information only.
The soundness of our approach is proved in the Abstract Interpretation
framework. We first construct a non-standard concrete semantics that gives
a formal meaning to unclean C constructs. Then, we abstract it to
derive a static analysis that is sound by construction.
This makes our design modular---it can be used with any underlying numerical 
domain, even a relational one such as octagons---and 
extensible---new abstractions based
on the same concrete semantics can be designed.
The abstraction
is currently limited to programs without unbounded dynamic memory
allocation or recursively.
We deliberately left out these features as they are generally forbidden in
critical software.
Our work was integrated within the \astree\ analyzer 
\cite{magic2} that checks for run-time errors in embedded
critical C code and provides tight variable bounds, in a few hours of
computation time.

\paragraph{Overview of the Paper.}
Sect.~\ref{motivsect} motivates our work by presenting a few realistic
code examples involving union types and complex pointer arithmetics;
they cannot be analyzed soundly without considering byte-level aliasing.
In Sect.~\ref{analsect}, we present our solution to this problem in
an intuitive way.
Sect.~\ref{formalsect} then formalizes our approach in the Abstract 
Interpretation framework.
Sect.~\ref{astreesect} presents preliminary experimental results obtained with
the \astree\ analyzer.
Sect.~\ref{relsect} presents related work.
Finally, Sect.~\ref{futuresect} discusses future work and 
Sect.~\ref{conclusionsect} concludes.

\section{The Need for a New Memory Domain}
\label{motivsect}

The simplest framework, when performing a value-analysis, is to consider
programs with a statically known set of variables, each having a {\em scalar\/}
type: real ({\em i.e.}, integer or floating-point) or pointer type. 
Such analyses can be lifted to cope with variables of 
{\em aggregate\/}\footnote{The
terms {\em real}, {\em scalar}, {\em aggregate\/} come from the C norm
\cite{cnorm}.} type---arrays and structures---by decomposing them
into collections of independent {\em cells\/} of scalar type.
Much literature has been devoted to the problems of abstracting numerical
invariants, performing pointer analysis, or summarizing aggregate variables 
into fewer cells---to cope with large arrays or dynamic memory allocation.
We are concerned here with the case where the basis hypothesis of these works
fail: the memory cannot be decomposed {\em a priori\/} into a set of
independent cells.
This happens in a language such as C that permits very low-level accesses to
the memory and the bit-representation of data.

\begin{figure}\begin{center}
{\tt\begin{tabular}{l}
struct msgA \{ int type; int a[2]; \};\\
struct msgB \{ int type; double x; \};\\
\\
union msg \{\\
\quad struct \{ int type; \} T;\\
\quad struct msgA A;\\
\quad struct msgB B;\\
\};\\
\\
void process(union msg *m) \{\\
\quad switch (m->T.type) \{\\
\quad case 0: \{\\
\quad\quad struct msgA* msga = \&(m->A);\\
\quad\quad int data = msga->a[0]+1;\\
\quad\quad /* {\em work on msga} */\\
\quad \}\\
\quad case 1: \{\\
\quad\quad struct msgB* msgb = \&(m->B);\\
\quad\quad /* {\em work on msgb} */\\
\}\\
\\
void read\_sensor\_4(unsigned* m) \{\\
\quad /* {\em put 4 bytes from sensors into m} */\\
\}\\
\\
void main(void) \{\\
\quad unsigned char buf[sizeof(union msg)];\\
\quad int i;\\
\quad for (i=0;i<sizeof(buf)/4;i++)\\
\quad \quad read\_sensor\_4((unsigned*)buf+i);\\
\quad process((union msg*)buf);\\
\}\\
\end{tabular}}
\caption{Message manipulation example illustrating the use of union types.}
\label{msgex}
\end{center}\end{figure}

\begin{figure}\begin{center}
{\tt\begin{tabular}{l}
void \\memcopy(void* dst, void* src, unsigned sz) \{\\
\quad unsigned char* s = (unsigned char*) src;\\
\quad unsigned char* d = (unsigned char*) dst;\\
\quad unsigned i;\\
\quad for (i=0;i<sz;i++) d[i] = s[i];\\
\}\\
\\
int get(unsigned char* buf) \{\\
\quad struct \{ int *p; $\cdots$\ \} S;\\
\quad memcopy(\&S,\,buf+16,\,sizeof(S));\\
\quad return *(S.p);\\
\}\\
\end{tabular}}
\caption{User-defined generic memory copy procedure.}
\label{memcpyex}
\end{center}\end{figure}

\begin{figure}\begin{center}
{\tt\begin{tabular}{l}
void \\memcopy(void* dst, void* src, unsigned sz) \{\\
\quad char* s = (char*) src;\\
\quad char* d = (char*) dst;\\
\quad for (;sz>=8;sz-=8,s+=8,d+=8)\\
\quad\quad *((double*)d) = *((double*)s);\\
\quad for (;sz!=0;sz{-}{-},s++,d++) *d = *s;\\
\}\\
\end{tabular}}
\caption{Alternate user-defined memory copy procedure.}
\label{memcpyex2}
\end{center}\end{figure}

\paragraph{Union Types.}
Union types declare fields that, unlike aggregate types,
share the same memory locations.
As a consequence, access paths to cells may be aliased.
Consider Fig.~\ref{msgex} implementing message objects
using a dynamic type tag {\tt type}.
In the {\tt process} function, {\tt m->\hi{}T.type},
{\tt m->\hi{}A.type} and {\tt m->\hi{}B.type} all
refer to the same cell containing
an {\tt int}.
It is perfectly legal to modify the cell using one access path
and read back its contents using another one \cite[\S 6.5.2.3.5]{cnorm}.
This kind of aliasing is quite benign as it does not prevent us from viewing
the memory as a collection of distinct cells---{\em e.g.}, using
{\em offsets\/} instead of access paths to denote cells.

A programmer may, however, disregard the value of {\tt type},
write into {\tt m->\hi{}A.a[0]} and read back {\tt m->\hi{}B.x},
thus mixing access paths referring to (partially) 
overlapping memory locations of different types.
Although such mixing is strongly discouraged by the 
C norm \cite{cnorm} 
and relies on unportable assumptions on structure layouts and value 
encodings, it is surprisingly often performed by programmers.
Consider the variable \mt{regs} modeling, in Fig.~\ref{emuex},
the register state of an Intel 8086 processor.
It is expected that, when modifying the word register
{\tt regs.w.ax}, its low- and hi-byte components
{\tt regs.b.ah} and {\tt regs.b.al} are updated and can
safely be read back.
Due to this {\em byte-level\/} aliasing, no partition of the memory into 
scalar cells exists.

\paragraph{Pointer Arithmetics.}
Pointer arithmetics encompasses array indexing.
For instance, given the following declaration:
\begin{center}
{\tt\begin{tabular}{l}
struct \{ int a[3]; int b; \} U, V;\\
\end{tabular}}
\end{center}
{\tt *(U.a+2)} is equivalent to {\tt U.a[2]}.
But pointer arithmetics also allows escaping from an array embedded within
a larger type, breaking standard out-of-bound array analyses.
For instance, {\tt *(U.a+3)} can safely be considered equivalent
to {\tt U.b} for most compilers.
No assumption can generally be made, however, on the relative position of
{\tt U} and {\tt V} in memory: {\tt U.a[4]} is considered a run-time error
and {\tt U.a+4} points to an unspecified location outside {\tt U}---generally 
not within {\tt V}.

\paragraph{Pointer Casts.}
Pointer casts allow considering any part
of the memory as having any type.
Consider the {\tt main} function in Fig.~\ref{msgex}.
It declares {\tt buf} as an array of {\tt unsigned char} 
but actually uses it both as
a reference to an {\tt unsigned int}
(when calling {\tt read\_\hi{}sensor\_\hi{}4}) and as
a message of type {\tt union msg} (when calling {\tt process}).
This achieves the same effect as a union type, except that the set of
possible cell layouts is no longer embedded within
the static type of the variable. It must be guessed dynamically.
An extreme illustration of this problem is given by the generic memory copy
functions {\tt memcopy} 
of Fig.~\ref{memcpyex} (a portable, one-byte-at-a-time version) and 
Fig.~\ref{memcpyex2} (an optimized version that copies by bunches of eight
bytes, inspired from actual PowerPC software).
There, the {\tt void*} type is used to achieved polymorphism. 
This effectively discards all type information that would hint at the structure
of the memory from {\tt src} to {\tt src+sz}-1.
Despite this lack of typing information, we must be able to copy multi-byte
cells from {\tt src} to {\tt dst} in a way consistent with their type.
In order to treat precisely the indirect addressing {\tt *(S.p)}
at the end of the {\tt get} function in Fig.~\ref{memcpyex}, it
is paramount to copy ``as-is'' the pointer value hidden at offset $16$
in {\tt buf}.
We refer the reader to Siff et al. \cite{repscast} for more 
examples of type casts used in real-life C programs.

\section{Overview of the Analysis}
\label{analsect}

In this section, we only try to present the gist of our analysis in an informal
way.
The next section will be devoted to its precise, mathematical definition.

\subsection{Assumptions}

\paragraph{Limitations.}
Our analysis computes, for each {\em control state}, an overapproximation
of the reachable memory states, where a control state is given by a program
point together with a call-stack.
For the sake of simplicity, we place ourselves in the context of a fully 
context-sensitive analysis on code without recursive procedure nor 
dynamically  memory allocation.
Our main hypothesis is that the set $\V_c$ of variables whose contents define 
the memory state---global and local variables from all stack 
frames---is a static function of the control state $c$ only.
In practice, it is valid when analyzing embedded C code (where {\tt malloc}
and recursion are prohibited) with a high level of precision (requiring
context-sensitivity).
However, we believe that these limitations may be overcome using 
{\em summarization\/} techniques which are orthogonal to our 
purpose---{\em e.g.}, heap abstraction as in \cite{tvla}, array summarization
\cite{weak}, or procedure summarization \cite{pt}.

\paragraph{Application Binary Interface.}
In order to achieve a high-level of description and discourage unportable
practices, the C norm \cite{cnorm} under-specifies many parts of the language.
In particular, the exact encoding of scalar types as well as the layout of
fields in structures are mostly left to the implementor. 
However, in order to ensure the interoperability of compiled programs, 
libraries, and operating systems, the precise representation of types is 
standardized in so-called implementation-specific
Application Binary Interfaces (or {\em ABI\/}) such as \cite{abi}.
Although it is possible to write fully portable, ABI-neutral C code, our
purpose here is the analysis of C programs that make explicit use
of architecture-dependent features---such as embedded programs that need to
be efficient and have a low-level access to the system.
Thus, our analysis is parameterized by ABI functions, such as
$\mt{sizeof}: \V_c\rightarrow \N$ that gives the byte-size of each variable.

\paragraph{Input Language.}
We suppose that each C function has been processed into a control-flow
graph where basic blocks are either assignment or guard instructions
involving only side-effect free expressions.
Moreover, using our knowledge of the ABI, all pointer arithmetics has been 
broken down to the byte-level. 
Except for the purpose of dereferencing, all
pointers can be assumed to be pointers to {\tt unsigned char}.
All memory reads and writes are performed through pointer dereferencing.
We assume that these involve only scalar types
({\em i.e.}, integers, floating-points, and pointers).
Likewise, field selection {\tt .} and {\tt ->} in {\tt struct} and 
{\tt union}, as well as array indexes {\tt [\,]} have been converted into
byte-level pointer arithmetics and dereferences of values of scalar type.
As these are usual static simplifications performed by most 
compilers and analyzer front-ends, we do not present them in more 
details.
Constructs that do not fit in this simplified framework
(such as function pointers or assignment of compound values)
will be dealt with in Sect.~\ref{formalsect}.

\paragraph{Numerical Analysis Parameter.}
Our analysis is parameterized by a standard numerical analysis.
Following the Abstract Interpretation framework \cite{ai}, we suppose that
it is given in the form of a {\em numerical abstract domain}, 
{\em i.e.}, an abstract representation of invariants together with
abstract {\em transfer functions\/} to mimic, in the abstract, 
the effect of instructions and control-flow joins.
In theory, such an analysis outputs an invariant $I_c$ on $\V_c$
at each program point $c$.
However, it supposes that variables are unaliased and have real 
type ({\em i.e.}, integer or floating-point), which is not the case for $\V_c$.
Thus, we do not use the numerical domain directly on $\V_c$ but on some
collection 
$\C_c$ of synthetic {\em cells\/} of real type.
We provide an abstraction of the memory layout that drives the numerical
analysis by dynamically managing $\C_c$, translating instructions over
$\V_c$ into instructions over $\C_c$, and taking care of
byte-level aliasing between cells in $\C_c$.

\subsection{Abstract Memory Layout}
\label{absmemsect}

Each variable $V$ is viewed as an unstructured sequence of 
${\tt sizeof}(V)$ contiguous bytes.
Its layout in $\C_c$ is initially empty. It will be populated with
possibly overlapping cells of scalar type as $V$ is accessed.
Abstracting a program instruction is done in three steps.
First, we {\em enrich\/} the layout by adding
all cells targeted by a dereference in the instruction.
Secondly, we evaluate, in the numerical domain.
the instruction where all dereferences have been replaced with cells.
Thirdly, we {\em remove\/} all cells invalidated by  alias-induced 
side-effects.
When a layout $\C_c$ is changed, the corresponding cells are created
or deleted in the numerical invariant $I_c$.

\begin{figure}\begin{center}
{\tt\begin{tabular}{l}
static union \{\\
\quad struct \{ uint8\ al,ah,bl,bh,\ldots\ \} b;\\
\quad struct \{ uint16\ ax,bx,\ldots\ \} w;\\
\} regs;\\
\\
regs.w.ax = X; {\em (1)}\\
if (!regs.b.ah)  {\em (2)} regs.b.bl = regs.b.al; {\em (3)}\\
else {\em (4)} regs.b.bh = regs.b.al; {\em (5)}\\
{\em (6)} regs.b.al = X; {\em (7)}\\
\end{tabular}}
\caption{Register state of an Intel 8086 processor and sample code to
manipulate it.
{\em (1)\/} to {\em (7)\/} 
represent program points of interest---see Fig.~\ref{emufig}.}
\label{emuex}
\end{center}\end{figure}

\begin{figure}\begin{center}
\includegraphics[width=4.8cm,height=9cm]{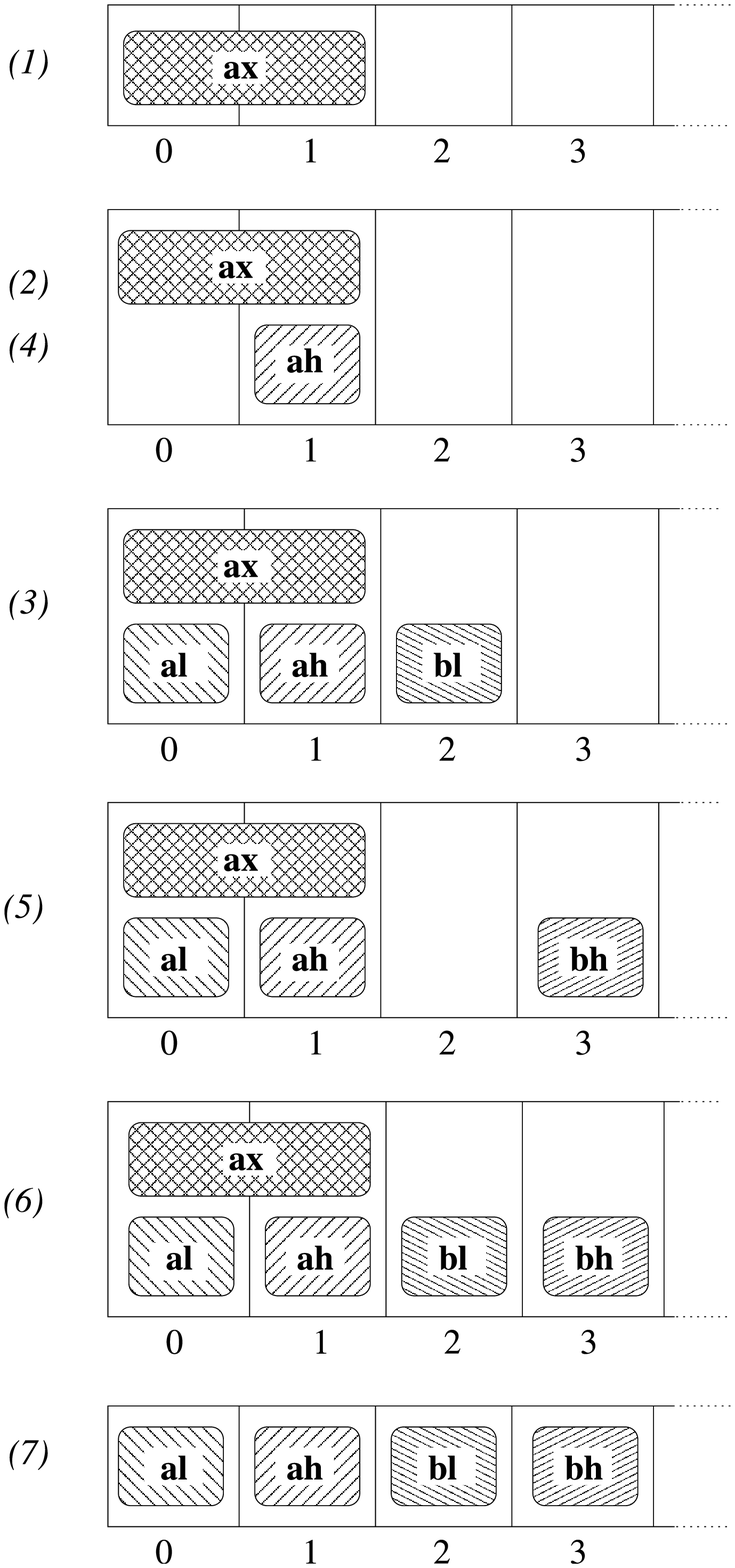}
\caption{Memory layouts $C_1$ to $C_7$ for the variable {\tt regs}
at program points {\em (1)\/} to {\em (7)\/}
when analyzing Fig.~\ref{emuex}.}
\label{emufig}
\end{center}\end{figure}

We illustrate this mechanism on the example of Fig.~\ref{emuex}.
Fig.~\ref{emufig} gives the abstract memory layouts $\C_1$ to $\C_7$ 
of the variable {\tt regs} at program points {\em (1)\/} to {\em (7)}.
\begin{itemize}
\item The assignment {\tt regs.w.ax = X} first creates a new cell, named
{\bf ax}, of type {\tt uint16}, occupying offsets $0$ and $1$ in the
variable {\tt regs}---see the top of Fig.~\ref{emufig}.
Supposing that {\tt X} corresponds to cell {\bf X}, the assignment is then
evaluated as ${\bf ax} \leftarrow {\bf X}$ in the numerical domain
to yield $I_1$.
\item Before executing the test {\tt !regs.b.ah}, 
the cell {\bf ah} of type {\tt uint8} is created at offset $1$ in {\tt regs}.
If the ABI tells us that the computer uses a little-endian 
byte-ordering, the cell can be initialized using the constraint
${\bf ah}={\bf ax}/256$ on $I_1$.
Finally, the test is executed by adding the constraint ${\bf ah}=0$.
When backed by a sufficiently powerful numerical domain, we may be
able to infer that ${\bf X}/256={\bf ax}/256={\bf ah}=0$, {\em i.e.},
${\bf X}\in[0,255]$, at {\em (2)}.
\item Let us now consider the control-flow join following the
conditional branches {\em (3)\/} and {\em (5)}.
As $\C_3\neq\C_5$, we must unify cell sets.
This is done by adding missing cells: ${\bf bh}$ is added to $\C_3$
and ${\bf bl}$ to $\C_5$.
As {\tt regs} is declared {\tt static}, its bytes are initialized to 
$0$---according to the C norm \cite{cnorm}.
Thus, we add the constraint ${\bf bh}=0$ to $I_3$ and ${\bf bl}=0$ to $I_5$.
The control-flow join is then performed safely in the numerical domain,
using the cell set $\C_6=\C_3\cup\C_5$.
\item
The last assignment, {\tt regs.b.al = X}, can be directly evaluated as
${\bf al}\leftarrow {\bf X}$ in the numerical domain because all the
cells involved exist.
However, modifying ${\bf al}$ also modifies ${\bf ax}$, a fact the numerical
domain is not aware of. 
We correct the invariant $I_7$ by deleting, after the assignment, all cells
that overlap the modified cells. Thus, ${\bf ax}\notin\C_7$.
Note that ${\bf ax}$ will be back when ${\tt regs.w.ax}$ 
is accessed next and 
its contents will be synthesized using fresh information from the overlapping 
cells ${\bf ah}$ and ${\bf al}$.
\item
In the presence of loops, we iterate the abstract computation until
it stabilizes.
Numerical domains usually use special joint-like binary operators 
$\widen^\s$, so called widenings \cite{ai}, to accelerate fixpoint computations.
As for control-flow joints, we first unify the cells layouts of the arguments,
and then apply $\widen^\s$ in the numerical domain.
\end{itemize}

\subsection{Pointer Abstraction}
\label{ptrsect}
Numerical domains can only abstract directly environments over 
cells of real type, not pointer type.
Thankfully, a pointer value can be viewed as a pair 
$(V,o)\in \V_c\times\N$, which represents the offset $o$, in bytes, 
from the beginning of the variable $V$.
We abstract each component independently: one cell of 
integer type is allocated in the numerical domain for each pointer cell
to represent its possible offsets
while we maintain, in the memory layout, a map associating a set of
base variables to each pointer cell.
Pointer arithmetics in expressions are straightforwardly translated into 
integer arithmetics on offsets and then fed to the numerical domain.
One benefit of this is that we are able to find relationship between pointer 
and integer values.
For instance, when using the polyhedron domain, we can infer that
{\tt s=src+sz} holds at the end of the memory copy procedure of 
Fig.~\ref{memcpyex2}.

\subsection{Intersection Semantics}
\label{intersect}
As it will become clear in Sect.~\ref{cellabssect}, we actually use an 
{\em intersection\/} semantics for overlapping cells.
Suppose, for instance, that the analysis of Fig.~\ref{emuex} found
some numerical invariant $I_c$ with respect to the layout 
$\C_c=\{{\bf ah},{\bf ax}\}$.
Then, taking byte-level aliasing into account,
${\tt regs.b.ah}=v$ is possible only if both
${\bf ax}$'s hi-byte and ${\bf ah}$ can be $v$:
$I_c$ actually stands for $I_c\wedge({\bf ah}={\bf ax}/256)$.
This semantics ensures that, when reading a cell's contents, it
is safe to ignore overlapping cells---we simply lose some constraints,
which is sound.
In practice, when there already exists a cell in $\C_c$ with the correct
type and offset---which is the most common case---we 
use it without looking at overlapping cells while, when
it needs to be created, we use existing overlapping cells to synthesize
good and safe initial values.
Dually, when writing a cell's contents, we must take care to update
{\em all\/} overlapping cells as they give constraints that were 
true before the assignment but are no longer valid.
In practice, we destroy such cells, which actually delegates the update 
to the next time the cell is created back by a read.

Another legitimate choice would have been a {\em union\/} semantics.
However, this would have made write cheap and read costly. We
favored the cheap reads of the intersection semantics.
Also, as we will see in Sect.~\ref{cellabssect}, the intersection semantics
has a very natural formalization.

For performance reasons in the numerical domain, we should avoid creating too 
many cells.
Our scheme keeps several redundant cells per memory byte.
Thankfully, redundancy is bounded by the low number of scalar types---13 as
shown in Fig.~\ref{langsyn}.
As cells are created lazily and destroyed often, there is few
long-lived redundancy in practice.
Moreover, by associating only one offset variable per pointer---and not one
for each basis variable the pointer can point to---we lose some precision but
avoid a potential quadratic blow-up.

\section{Formalization of the Analysis}
\label{formalsect}

\subsection{Abstract Interpretation}

We formalize our analysis in the Abstract Interpretation framework, 
a general theory
of the approximation of program semantics introduced by Cousot 
and Cousot in \cite{ai}.
It allows the systematic design of static analyses 
with various levels of precision.
The gist of the method is first to design a {\em concrete semantics}, the
most precise mathematical expression of program behaviors.
This step emphases on expressibility only and generally results in a non
computable semantics.
Then, sound abstractions are performed and composed until a computable 
semantics is derived from the concrete one.
This results in an abstract interpreter that can be run without user
intervention, terminates on all programs, and is, by construction, sound
with respects to the concrete semantics.
It is, however, often incomplete.
Abstractions should be carefully chosen based on the class of properties
to be checked, the class of programs analyzed, and the amount of resources
to be invested in the static analysis.

There exists a large library of {\em abstract domains\/} that provide
ready-to-use abstract computation algorithms.
Formally, given a concrete universe $\D$ where the concrete semantics is 
formalized, an abstract domain is given by:
\begin{itemize}
\item a set $\D^\s$ of computer-representable abstract properties,
\item a concretization function $\gamma:\D^\s\rightarrow\P(\D)$ assigning
a meaning to each abstract property,
\item a computable partial order $\sqsubseteq^\s$ on $\D^\s$ such that
$\gamma$ is monotonic: 
$X^\s\sqsubseteq^\s Y^\s\Longrightarrow \gamma(X^\s)\subseteq\gamma(Y^\s)$;
it models the relative precision of abstract elements and enables
fixpoint abstractions through iteration schemes---potentially
accelerated using special extrapolation operators $\widen^\s$,
\item for each $n-$ary
semantical operator $F:\D^n\rightarrow\P(\D)$, an abstract, computable
version $F^\s:\D^\s{}^n\rightarrow\D^\s$ that is sound, {\em i.e.},
$\forall X^\s_i\in\D^\s,$ $\forall X_i\in\gamma(X^\s_i),$
$F(X_1,\ldots,X_n)\subseteq (\gamma\circ F^\s)(X^\s_1,\ldots,X^\s_n)$.
\end{itemize}

We refer the reader to \cite{ai,ai2} for more informations on
the theory of Abstract Interpretation and its applications.

\subsection{Language}

\begin{figure}\begin{center}\begin{tabular}{l}
\hspace*{-0.25cm}
$\begin{array}{lcl}
\mi{int-sign} &::= &\mt{unsigned}\ |\ \mt{signed}\\
\mi{int-type} &::= &\mt{char}\ |\ \mt{short}\ |\ \mt{int}\ |\ \mt{long}\ 
|\ \mt{long long}\\
\mi{float-type} &::= &\mt{float}\ |\ \mt{double}\ |\ \mt{long double}\\
\mi{real-type} &::= & \mi{int-sign int-type}\ |\ \mi{float-type}\\
\mi{scalar-type}\hspace*{-0.2cm} & ::= & \mi{real-type}\ |\ \mt{ptr}\\
\mi{type} &::= & 
    \mi{scalar-type}\\
&|& \mi{type}\mt{[}n\mt{]} \quad n\in\N\\
&|& \mt{struct}\ \mt{\{}\ \Id_1:\mi{type},\ldots,\Id_n:\mi{type}\ \mt{\}}\\
&|& \mt{union}\ \mt{\{}\ \Id_1:\mi{type},\ldots,\Id_n:\mi{type}\ \mt{\}}\\
\end{array}$
\\\\
$\begin{array}{lcll}
\mi{expr} &::=&  
    \mi{cst} & \mi{cst}\in\R\\
&|& \mt{\&}V & V\in\V_c\cup\F\\
&|& \diamond\,\mi{expr} & \diamond\in\{-,\sim,!\}\\
&|& \mi{expr}\diamond\mi{expr} & \diamond\in\{+,\leq,\mt{\&},\mt{||},\ldots\}\\
&|& \mt{*}_\tau\,\mi{expr} & \tau\in\mi{scalar-type}\\
&|& \mt{(}\tau\mt{)}\,\mi{expr} & \tau\in\mi{scalar-type}\\
\end{array}$
\\\\
$\begin{array}{lcll}
\mi{inst} &::= & 
    \mt{*}_\tau\,\mi{expr} \leftarrow \mi{expr} & \tau\in\mi{scalar-type}\\
&|& \mt{*}_\tau\,\mi{expr} \leftarrow \mt{*}_\tau\,\mi{expr} & \tau\in\mi{type}\\
&|& \mi{expr} == 0\; ?\\
\end{array}$
\end{tabular}
\caption{Language Syntax.}
\label{langsyn}
\end{center}\end{figure}

We suppose that the program has been preprocessed into the simple
language of Fig.~\ref{langsyn}.
Each type denotes not only a set of possible values, but also their
bit-representation in memory.
We assume that all pointers use the same representation, and so, 
use a single type denoted by {\tt ptr}.
In expressions, $\V_c$ and $\F$ denote respectively the set of variables
at control point $c$ and functions.
We have distinguished two kinds of assignments: assignments of 
expressions of scalar type, and {\em copy\/} assignments of arbitrary
data structures.

\subsection{Concrete Memory Domain $\D_M$}

We now introduce a non-standard, low-level semantics $\D_M$ 
that gives a meaning to the programs of Sect.~\ref{motivsect}.

\subsubsection{Concrete Memory Representation}
\label{datasect}

\paragraph{Values.}
Let us denote by $\Val_\tau$ the set of values of scalar type $\tau$.
For real types, $\Val_\tau$ is a finite subset of $\R$.
Pointer values range in the following set:
$$\!\!\begin{array}{l}
\Val_\mt{ptr} \deq 
\{\,(V,i)\,|\,V\in\V_c\cup\F,\,0\leq i\leq\mt{sizeof}(V)\,\}
\cup\{\nullptr,\err\}\\
\end{array}\!\!\!$$
where $\nullptr$ is the {\tt NULL} pointer while $\err$ represents
all erroneous pointer values.
Valid pointers pointers are (base,offset) pairs.
Following the C norm, data pointers can point one byte past the end of a 
variable.
To treat function pointers the same way as data pointers, it is
sufficient to extend $\mt{sizeof}$ so that $\mt{sizeof}(V)\deq0$ when $V\in\F$:
valid functions pointers always have a null offset.

\paragraph{Memory State.}
We decompose the memory into a collection
$\B$ of untyped {\em byte locations\/}:
$$\B(\V_c)\deq \{\;(V,i)\;|\;V\in\V_c,\;0\leq i<\mt{sizeof}(V)\;\}$$
The set of values a byte can hold is defined as the following set $\Val$
of triples:
$$\Val\deq
\{(\tau,b,v)|\tau\in\mi{scalar-type},0\leq b<\mt{sizeof}(\tau),
v\in\Val_\tau\}$$
where $(\tau,b,v)$ represents symbolically the $b-$th byte of the 
representation of the value $v$ of scalar type $\tau$.
A concrete memory state associates a byte value to each byte 
location: $\D_M(\V_c)\deq \B(\V_c)\rightarrow \Val$.
Note that, in actual computers, the memory maps byte locations to
numbers within $[0,255]$.
Our memory representation is sightly higher-level as it abstracts away
the encoding from $\Val$ to $[0,255]$.
For instance, the base variable of pointers is kept
symbolic so that our semantics is independent from the absolute
address chosen for the variables by memory allocation services.

\begin{figure}\begin{center}
$\begin{array}{l}
\phi_{\tau}\langle(\tau_0,b_0,v_0),\ldots \rangle\deq \{v\}
\quad\text{if}\;\forall k,\,v_k=v,\,\tau_k=\tau,\,b_k=k
\\
\phi_{\mt{unsigned char}}\langle(\tau,b,v)\rangle\deq\\
\qquad
\left\{\begin{array}{ll}
\{0\} & \text{ if }\tau=\mt{ptr}\text{ and }v=\nullptr\\
\{v/(256^b)\text{ mod }256\} & \text{ if }\tau\in\mi{int-type}\\
{}[0,255] & \text{ otherwise}\\
\end{array}\right.
\\
\phi_{\mt{unsigned}\ t}\langle x_0,\ldots\rangle\deq
\{\begin{array}[t]{l}
  \sum_k 2^{256\times k}\times y_k\;|\\
   y_k\in\phi_\mt{unsigned char}\langle x_k\rangle\;\}
\end{array}
\\
\phi_{\mt{signed}\ t}\langle x\rangle\deq
\{\begin{array}[t]{l}w\;|\;
w+2^{\mt{sizeof}(t)}\Z\cap\phi_{\mt{unsigned}\ t}\langle x\rangle\neq\emptyset,\\
w\in[-2^{\mt{sizeof}(t)-1}-1,2^{\mt{sizeof}(t)-1}]\;\}
\end{array}
\\
\phi_\mt{ptr}\langle x\rangle \deq 
\left\{\begin{array}{ll}
\{\nullptr\} & \text{if $\phi_\mt{unsigned long}\langle x\rangle=0$}\\
\Val_\mt{ptr} & \text{otherwise}
\end{array}\right.
\\
\text{in all other cases, }\phi_\tau\langle x\rangle \deq  \Val_\tau\\
\end{array}$
\caption{Value recomposition function example.}
\label{recompfig}
\end{center}\end{figure}

\paragraph{Value Recomposition.}
Due to pointer casts, a
sequence of bytes may be dereferenced as a value of any type.
Thus, we now suppose that we are given a family of functions $\phi_\tau$ that 
construct all the values of type $\tau\in\mi{scalar-type}$ 
corresponding to a given byte sequence:
$\phi_\tau:\Val^{\mt{sizeof}(\tau)}\rightarrow \P(\Val_\tau)$.
Note that, to allow a conservative modeling of casting, the functions may 
output several values.
The exact definition of $\phi$ is highly dependent upon the ABI. 
We provide, in Fig.~\ref{recompfig}, an example definition valid for Intel 
x86 processors.
It embeds useful information, such as the fact that \nullptr\ 
is always represented as the integer $0$, or that integers are represented using
two's complement arithmetics and little endian byte ordering---it models
precisely the \mt{regs} variable in Fig.~\ref{emuex}.
However, when the value depends upon information abstracted away by our 
semantics ({\em e.g.}, non-\nullptr\ pointers cannot be converted to
integers without knowing the absolute address of variables) 
or when we are not interested in the precise 
behavior of a particular construction ({\em e.g.}, reading the 
binary representation of floating-point 
values) we use a conservative definition:
$\phi_\tau\langle x \rangle=\Val_\tau$.
If need be, these cases can be refined.
Dually, we may trade precision for generality---{\em e.g.}, drop our
assumption on the byte ordering of integers.

\subsubsection{Concrete Semantics}

\paragraph{Expression Semantics.}
The concrete semantics $\lb{e}:\D_M(\V_c)\rightarrow \P(\Val_\tau)$
of an expression $e$ of type $\tau\in\mi{scalar-type}$ 
associates a set of values to a memory state.
Most of its definition can be readily extracted from the C norm
\cite{cnorm} and the IEEE 754-1985 norm \cite{ieee754}.
We present here only the part related to our non-standard definition of the 
memory.
It corresponds to the semantics of pointers and dereferences:
\begin{itemize}
\item 
$\lb{\mt{\&}V}(M)\deq\{\,(V,0)\,\}$
\item
$\lb{e+e'}(M)  \deq \{\begin{array}[t]{l}
(V,i+j)\;|\;
0\leq i+j\leq \mt{sizeof}(V),\;\\
(V,i)\in\lb{e}(M),\;j\in\lb{e'}(M)\;\}
\end{array}$
\item
$\lb{e-e'}(M)  \deq 
\{\begin{array}[t]{l}i-j\;|\\
(V,i)\in\lb{e}(M),\;(V,j)\in\lb{e'}(M)\;\}\end{array}$
\item
$\lb{\mt{(ptr)}e}(M) \deq
\left\{\begin{array}{ll}
\{\nullptr\} & \text{if $\lb{e}(M)\subseteq\{0,\nullptr\}$}\\
\Val_\mt{ptr} & \text{otherwise}
\end{array}\right.$
\item
$\lb{\mt{*}_\tau\,e}(M) \deq\\
\qquad
\cup\;\{\begin{array}[t]{l}
\phi_\tau\langle M(V,i),\ldots,M(V,i+\mt{sizeof}(\tau)-1)\rangle\;\;|
\\
(V,i)\in\lb{e}(M),\;i+\mt{sizeof}(\tau)\leq\mt{sizeof}(V),
\\
i\equiv 0\;[\mt{alignof}(\tau)]\;\}\end{array}
$
\end{itemize}
As before, the non-determinism allows a lose but sound modeling of concrete
actions.
Erroneous computations (such as overflows in pointer arithmetics and
out-of-bound or misaligned pointers in dereferences)
halt the program, and so,
do not contribute to the set of accessible states.

\paragraph{Instruction Semantics.}
The semantics $\lc{i}:\D_M(\V_c)\rightarrow\P(\D_M(\V_c))$ of an instruction $i$
maps a memory state before the instruction to a set of possible memory states
after the instruction.
It is defined as follows:
\begin{itemize}
\item tests filter out environments that cannot satisfy the test:
$$
\lc{e==0\;?}(M) \deq \left\{\begin{array}{ll}
\{M\} & \text{if $0\in\lb{e}(M)$}\\
\emptyset & \text{otherwise}
\end{array}\right.$$
\item copy assignments perform a byte-per-byte copy:
$$\hspace*{-0.5cm}
\begin{array}{l}
\lc{\mt{*}_\tau\,e\leftarrow \mt{*}_\tau\,f}(M)\deq \{\\
\begin{array}[t]{l}
M[(V,i)\mapsto M(W,j),\ldots,(V,i+n)\mapsto M(W,j+n)]\;
\;|\;\\
(V,i)\in\lb{e}(M),\,
(W,j)\in\lb{f}(M),\,
n=\mt{sizeof}(\tau)-1,\\
i+n<\mt{sizeof}(V),\;
j+n<\mt{sizeof}(W)
\;\}\end{array}
\end{array}
$$
\item regular assignments evaluate the right-hand expression
and store its byte components into the memory:
$$
\begin{array}{l}
\lc{\mt{*}_\tau\,e\leftarrow f}(M)\deq\\
\quad
\{\begin{array}[t]{l}
M[(V,i)\mapsto (\tau,0,v),\ldots,(V,i+n)\mapsto (\tau,n,v)]\;
\;|\;\\
(V,i)\in\lb{e}(M),\;
v\in\lb{f}(M),\;\\
n=\mt{sizeof}(\tau)-1,\;
i+n<\mt{sizeof}(V)
\;\}\end{array}
\end{array}
$$
\end{itemize}
Note how the value conversion $\phi$  due to pointer casts only occurs at
memory reads, {\em i.e.}, in a lazy way, so as to reduce the precision loss.
Most of the time, we fall in the first case of Fig.~\ref{recompfig}:
we read back a byte sequence corresponding to a value $v$
stored by a previous assignment of matching type;
$\phi$ returns the singleton $\{v\}$ and there is no loss of precision.

When $\tau$ is scalar, $\mt{*}_\tau\,e\leftarrow \mt{*}_\tau\,e'$ 
can be considered as either type of assignments,
but the copy assignment form is more precise because it
avoids interpreting the memory contents via $\phi$.
This allows the precise modeling of the polymorphic memory
copy functions of Figs.~\ref{memcpyex}--\ref{memcpyex2} as byte-per-byte
copies.

\paragraph{Variable Creation and Destruction.}
When creating a new (zero-initialized) variable $V$ of type $\tau$, new byte 
locations initialized to the value $(\mt{unsigned char},0,0)$ are added to $\B$.
However, deleting a variable $V$ from a memory state $M$ is more complex.
We must not only remove some byte locations from $\B$, but also invalidate
pointers to $V$ in the remaining locations, which gives the following 
memory state:
$$
(W,i) \mapsto
\left\{\begin{array}{ll}
(\mt{ptr},j,\omega) & \text{if $M(W,i)=(\mt{ptr},j,(V,\cdot))$}\\
M(W,i) & \text{otherwise}
\end{array}\right.
$$

\subsection{Memory Abstractions}

We now present computable abstractions of the concrete memory domain.
We are able to retrieve, in Sect.~\ref{cellabssect},
the analysis presented in Sect.~\ref{analsect}, in a sound and formal way.
We also present, in Sect.~\ref{equalsect},
a {\em memory equality\/} abstraction to improve its precision 
in the presence of copy assignments.

\subsubsection{Scalar Value Abstraction}
We suppose that we are given a numerical abstract domain
$\D^\s_\R(N)$ able to abstract environments over a set $N$ of cells with
real type.
That is, its concretization $\gamma_\R$ lives in $\D^\s_\R(N)\rightarrow\R^N$
and it features assignment and test transfer functions on expressions
involving only real-valued constants, cells in $N$, and arithmetic operators.
We refer the reader to \cite{ai,mine:float} 
for example definitions, including support for relational invariants and 
floating-point arithmetics.

Our first task is to add support for pointer values $\Val_\mt{ptr}$ to 
$\D^\s_\R(N)$.
As explained in Sect.~\ref{ptrsect}, the base component of a pointer is 
abstracted  as a set of variables or functions while
its offset is assigned a dimension in the numerical domain.
Given a collection $\C$ of cells of scalar type, the enhanced domain 
$\D^\s_\Val(\C)$ is constructed as follows:
$$\D^\s_\Val(\C)\deq
\D^\s_\R(\C)\;\times\;
(\C_\mt{ptr}\rightarrow \P(\V_c\cup\{\err,\nullptr\}))$$
where $\C_\mt{ptr}$ is the subset of $\C$ with pointer type.
A pair $(N,P)\in\D^\s_\Val(\C)$ 
represents the set $\gamma_\Val((N,P))$ of environments
$\rho:\C\rightarrow \cup_\tau (\Val_\tau)$ such that, for 
some $\sigma\in\gamma_\R(N)$:
if $V$ has real type, then $\rho(V)=\sigma(V)$;
if $V$ is a pointer,
then either $\rho(V)\in P(V)\cap\{\omega,\nullptr\}$ or
$\rho(V)\in (P(V)\setminus\{\omega,\nullptr\})\times\{\sigma(V)\}$.

At the level of $\D^\s_\Val$, we accept the same expressions
as in $\D^\s_\R$ with the
addition of pointer arithmetics---excluding pointer dereferencing.
As pointer arithmetics has been broken down to the byte level,
we can feed any instruction directly to $\D^\s_\R$ and obtain
its effect on the offset information.
The effect on pointer bases is derived by structural induction on expressions.
For instance, if \mt{p}, \mt{q} and \mt{i}
are respectively two pointers and an integer
variable, then the assignment 
$\lc{\mt{q}\leftarrow \mt{p}+\mt{i}}^\s_\Val(N,P)$ 
in $\D^\s_\Val$ will return the abstract pair 
$(\lc{\mt{q}\leftarrow \mt{p}+\mt{i}}^\s_\R(N),\,P[\mt{q}\mapsto P(\mt{p})])$ 
stating that
$\mt{q}$ now points to the same base variables as $\mt{p}$, and its offset is
that of $\mt{p}$ plus $\mt{i}$.
The binary abstract operators---such as union
$\cup^\s_\Val$ and ordering $\sqsubseteq^\s_\Val$---are defined point-wisely.
These are quite unoriginal, and so, we do not detail them further.

\subsubsection{Offset Abstraction}
\label{pointersect}

In practice, $\D^\s_\R$ is not a single numerical domain but a {\em reduced 
product\/} of several domains specifically chosen to fit the kinds 
of invariants 
found in an application domain---in our case, reactive control-command 
software, this includes plain intervals \cite{ai}, relational octagons 
\cite{mine:oct}, and domain-specific filter domains \cite{filters}.
Now that we rely on $\D^\s_\R$ to also abstract pointer offsets,
new kinds of numerical invariants are needed and we must enrich
our product.
An important property to infer is pointer alignment, such as
$\mt{p}\equiv 0\;[4]$ when $\mt{p}$
is used to access elements of byte-size $4$ in
an array.
For this, we use the {\em simple congruence domain\/} \cite{cong,repsx86}.

Although the combination of intervals and congruences seems sufficient
in most cases, preliminary experiments suggest the need to infer invariants 
of the more general form $\mt{p}\in\sum_i [a_i,b_i]\times c_i$
to represent, {\em e.g.}, slices in multi-dimensional arrays.
No such domain exists; its construction is left as future work.
Alternate ideas include using the reduced product of linear equalities
and intervals, as done by Venet \cite{venet}.

\subsubsection{Cell-Based Memory Abstraction $\D^{\s \Val}_M$}
\label{cellabssect}

\paragraph{Cell Universe.}
In order to use the value domain $\D^\s_\Val$, we need to map memory bytes 
in $\D_M(\V_c)$ to cells of scalar type.
Given $\rho\in\D_M(\V_c)$, for each binding $\rho(V,i)=(\tau,b,v)$, we must 
consider a cell of type $\tau$ at offset $i-b$ in variable $V$, with value
$v\in\Val_\tau$.
We define the following {\em cell universe\/}:
$$\C_\mr{all}(\V_c)\deq \{\,(V,i,\tau)\,|\hspace*{-0.1cm}
\begin{array}[t]{l}V\in\V_c,\,\tau\in\mi{scalar-type},\\
i\geq 0,\,i+\mt{sizeof}(\tau)\leq\mt{sizeof}(V)\,\}\end{array}$$
where $(V,i,\tau)$ corresponds to a cell of type $\tau$ starting at offset
$i$ in variable $V$.
It models bytes at locations $(V,i+b)$ for all $b$ in 
$[0,\,\mt{sizeof}(\tau)-1]$.
We will say that two cells {\em overlap\/} when the byte locations they
model overlap.
When extracting cells from a concrete state, we can
encounter overlapping cells---{\em e.g.}, 
$({\tt regs},0,{\tt uint16})$ and $({\tt regs},1,{\tt uint8})$
at program point {\em (2)\/} in  Fig.~\ref{emuex}.
As an abstract memory state is supposed to represent a {\em set\/} of concrete 
states, we must consider {\em a fortiori\/} overlapping cells to accurately 
model all possible memory structures.

\paragraph{Abstract States.}
An abstract memory state, in $\D^{\s \Val}_M$,
is given by a subset $\C$ of the cell universe, 
together with an abstract element in $\D^\s_\Val(\C)$ giving the cell contents:
$$\D^{\s \Val}_M(\V_c)\deq
\{\;(\C,X)\;|\;\C\subseteq\C_\mr{all}(\V_c),\;X\in\D^\s_\Val(\C)\;\}$$
A pair represents the following set of memory states:
$$\begin{array}{l}
\gamma^\Val_M(\{(V_1,i_1,\tau_1),\ldots,(V_n,i_n,\tau_n),\,X)\deq 
\{\;\rho\in\D_M(\V_c)\;|\;\\
\quad \forall x_1\in\phi_{\tau_1}
\langle \rho(V_1,i_1),\ldots,(V_1,i_1+\mt{sizeof}(\tau_1)-1) \rangle,\\
\quad\;\;\vdots\\
\quad \forall x_n\in\phi_{\tau_n}
\langle \rho(V_n,i_n),\ldots,(V_n,i_n+\mt{sizeof}(\tau_n)-1) \rangle,\\
\quad (x_1,\ldots,x_n)\in\gamma_\Val(X)\;\}
\end{array}$$
Note the universal quantifiers which mean that, when two cells from $\C$
overlap at a byte location $(V,i)$, $\gamma^\Val_M(\C,X)$ selects only concrete 
environments whose byte values at $(V,i)$ are compatible with {\em both\/}
cell values from $\gamma_\Val(X)$.
Hence the term {\em intersection semantics\/} used in Sect.~\ref{intersect}.
Moreover, when a byte location is not covered by any cell in $\C$, it can take
any value in the concrete world.

\paragraph{Cell Realization.}
It would be conceptually simpler to always consider $\C=\C_\mr{all}$, but
quite costly as the time and memory complexity of $\D^\s_\Val$ 
depends directly on the size of $\C$.
Thus, $\C$ is chosen dynamically.
As $\gamma^\Val_M$ has universal quantifiers, it is always safe to remove
any cell $c$ from $\C$:
$\gamma^\Val_M(\C,X)\subseteq
\gamma^\Val_M(\C\setminus\{c\},X_{|_{\C\setminus\{c\}}})$.
Adding a new cell $c$ is more complex:
we must initialize its value according to existing cells overlapping $c$ so as 
not to forget any concrete state.
We call this operation {\em cell realization}.
First, the cell is created and initialized to $\Val_\tau$, which
is sound.
Then, the value is refined by scanning the set of overlapping cells for
certain patterns and applying tests transfer functions in $\D^\s_\V$ 
accordingly.
For instance, when trying to add the cell ${\bf ah}$ in
the cell set $\C_1$ of Fig.~\ref{emufig}, one finds the overlapping cell
${\bf ax}$.
According to the $\phi_\mt{unsigned char}$ function of Fig.~\ref{recompfig},
we can apply the transfer function $\lc{{\bf ax}/256-{\bf ah}==0\,?}^\s_\R$.
Note that, if $\D^\s_\R$ contains relational domains, the relationship between
the realized and the overlapping cells will be kept.
For instance, if $\D^\s_\R$ is able to represent the invariant
${\bf ah}={\bf ax}/256$, then ,whenever we learn something new on the 
value of one cell, it will be immediately reflected on the other one.

\paragraph{Abstract Operators.}
Assignments and tests are transformed by replacing dereferences
with cell sets, and then fed to the underlying value domain.
Given a sub-expression $\mt{*}_\tau\,e$, where $e$ is dereference-free,
$e$ is first evaluated in $\D^\s_\Val$ 
which returns the set $S$ of byte locations it can point to.
All cells $\C'=\{\;(V,i,\tau),\;|\;(V,i)\in S\;\}$ are then realized in the
current abstract state---if not already there.
The resolution continues with the enriched abstract state
for the expression where $\mt{*}_\tau\,e$ has been replaced with the cell set 
$\C'$.
Tests can be directly executed in $\D^\s_\Val$ on the resulting expressions.
Assignments are a little more complex because they involve memory writes.
Given an assigned cell $c$, we first realize $c$, then execute the
assignment in $\D^\s_\Val$, and finally remove {\em all\/} 
cells overlapping $c$.
Note that a dereference may resolve in more than one cell, $|\C'|>1$, which
results in {\em weak updates\/} in $\D^\s_\Val$.
We now define the abstraction $\circ^{\s \Val}_M$ of a binary operator $\circ$.
Given the states $S_1=(\C_1,X_1)$ and $S_2=(\C_2,X_2)$, we first unify
the cell sets using realization to obtain two states
$S'_1=(\C_1\cup \C_2,X'_1)$ and $S'_2=(\C_1\cup \C_2,X'_2)$.
We then apply the binary operator on the underlying value domain and get
$S_1\circ^{\s \Val}_M S_2=(\C_1\cup \C_2,X'_1\circ^\s_\Val X'_2)$.
This is sound with respect to overlapping cells.
However, because overlapping cells have an intersection semantics, we may
lose some precision on $\cup^{\s \Val}_M$---informally, we over-approximate
$(a\cap b)\cup(c\cap d)$ as $(a\cup c)\cap(b\cup d)$.
The widening $\widen^{\s \Val}_M$ stabilizes invariants by first stabilizing the
cell set---which is an increasing subset of the finite set 
$\C_\mt{all}$---and then relies on the underlying widening $\widen^\s_\Val$.
The abstract order $\sqsubseteq^{\s \Val}_M$ is defined as
$\sqsubseteq^\s_\Val$ after cell sets have been unified to $\C_1\cup \C_2$.

There are strong similarities between the abstract cell realization
and the concrete value recomposition $\phi$.
Both are used, in a lazy way, to reconstruct information when the type of
a dereference mismatches that of the currently stored value.
Both are defined according to an ABI and the level of modeling required by the 
user.
Both may result in a loss of precision.
Thus, once a cell is realized, we try to keep it around as
long as possible ({\em i.e.}, until it is invalidated by a memory write).

\subsubsection{Memory Equality Predicate Domain $\D^{\s\mr{Eq}}_M$}
\label{equalsect}

When analyzing generic memory copy functions, $\D^{\s\Val}_M$ sometimes lacks the
required precision.
Consider, for instance, calling the function {\tt memcopy(\&a,\&b,4)} 
from Fig.~\ref{memcpyex}, {\tt a} and {\tt b} being 4-byte integers.
Although it is equivalent to the plain assignment {\tt a=b}, it is carried-out
one byte at a time. 
 $\D^{\s\Val}_M$ will first realize individual bytes in {\tt b} as 
{\tt char} cells, copy them into {\tt a} and, the first time {\tt a} is read, 
realize back the four {\tt char} cells as a single integer cell.
Because each realization may result in some loss of precision,
the inferred value set for the cell $({\tt a},0,{\tt int})$ 
may be much larger than that of $({\tt b},0,{\tt int})$.

In order to solve this problem, we introduce a specific abstraction
$\D^{\s\mr{Eq}}_M$ of $\D_M$ that tracks equalities between byte values in
a symbolic way:
$$\D^{\s\mr{Eq}}_M(\V_c)\deq \V_c\rightarrow 
((\N\times\V_c\times\N\times\N)\cup\{\top^{\s\mr{Eq}}\})$$
where a binding $V\mapsto(s,W,d,l)$ means that the $l$ bytes starting at
location $(V,s)$ are equal to those starting at location $(W,d)$, while
$\top^{\s\mr{Eq}}$ means ``no information:''
$$
\gamma^\mr{Eq}_M(\epsilon)\deq \{
\begin{array}[t]{l}\rho\in\D_M(\V_c)\;|\;
\forall V\in\V_c,\;
\epsilon(V)=(s,W,d,l)\\\Longrightarrow
\forall 0\leq i<l,\;\rho(V,s+i)=\rho(W,d+i)\;\}
\end{array}
$$
Note that only one predicate is kept per variable, and the parameters 
$(s,W,d,l)$ are bound to concrete values.
This ensures efficient transfer functions but requires memory copy loops to
be fully unrolled.
(We could benefit from more complex predicate abstraction schemes 
to overcome this  restriction---{\em e.g.}, use \cite{paramabs} to keep 
$(s,d,l)$ symbolic and relate their value in $\D^\s_\R(N)$.
This was not required in our experience as the codes we analyze only copy
small structures.)

Among instructions, only copy assignments are treated precisely:
tests are safely ignored while other assignments are dealt with by
removing bindings involving the destination---{\em i.e.}, setting them to 
$\top^{\s\mr{Eq}}$.
Suppose that $\epsilon(V)=(s,W,d,l)$ and we copy $l'$ bytes from $(V,s')$ to 
$(W',d')$; several cases arise.
When $W=W'$, $s-d=s'-d'$ and $s'\in[s,s+l]$, we copy bytes at the end of
equal zones.
We thus grow the zones by setting $\epsilon(V)=(s,W,d,\max(l,l'-s'+s))$.
The case is similar when bytes are copied at the start of zones:
$W=W'$, $s-d=s'-d'$ and $s\in[s',s'+l']$.
In all other cases, the former binding is useless and we replace it by a new
one $\epsilon(V)=(s',W',d',l')$.
As $W'$ is modified, we must also, in all cases, remove any other
binding involving $W'$.
We say that $\epsilon_1\sqsubseteq^{\s\mr{Eq}}_M\epsilon_2$ 
whenever, for every $V$,
either $\epsilon_2(V)=\top^{\s\mr{Eq}}$ or $\epsilon_1(V)$ corresponds to a
sub-range of $\epsilon_2(V)$.
This order has a least upper bound, which serves to define the abstract
union, but no greatest lower bound.
As $\D^{\s\mr{Eq}}_M$ has a finite height,
no widening is necessary to help the iterates converge.

We perform a partially reduced product between $\D^{\s \Val}_M$ and 
$\D^{\s\mr{Eq}}_M$.
All abstract operations are performed in parallel.
In addition, we propagate information from $\D^{\s\mr{Eq}}_M$
to $\D^{\s \Val}_M$ after each copy assignment.
For each cell $(V,o,\tau)$, if we just discovered that
$\epsilon(V)=(s,W,d,l)$ and $[o,o+\mt{sizeof}(\tau)]\subseteq[s,s+l]$, 
then we realize the cell 
$(W,o-s+d,\tau)$ and perform the assignment
${\tt *}_\tau (\mt{\&}W+o-s+d) \leftarrow {\tt *}_\tau (\mt{\&}V+o)$
in $\D^{\s \Val}_M$.
In our example, {\tt memcopy(\&a,\&b,4)}, we would generate the
assignment {\tt a$\leftarrow$b} just after copying the $4-$th byte.
Thus, the value for the cell $({\tt a},0,{\tt int})$ 
is precisely that of $({\tt b},0,{\tt int})$ and
no longer need to be realized from the value of {\tt char} cells.

\section{Experiments}
\label{astreesect}

\subsection{Presentation of the \astree\ Analyzer}

\paragraph{Scope.}
The goal of \astree\ is to detect statically all
run-time errors in embedded reactive software written in C.
Run-time errors include integer and floating-point arithmetics overflows,
divisions by zero and array out-of-bound accesses.
To achieve this goal, \astree\ performs an abstract reachability analysis and 
computes the set of values each variable can take, considering
all program executions in all possible environments.
To be efficient, it performs many sound but incomplete abstractions.
As a consequence, it always finds {\em all\/} run-time errors but
may report spurious alarms.
Its abstractions are tuned towards specific classes of programs in order to 
achieve {\em zero false alarms\/} in practice, within reasonable time.
Indeed, \cite{magic2} reports its success in proving automatically
the absence of run-time errors in real industrial code of several hundred 
thousand lines, in a few hours.

\paragraph{Architecture.}
\astree\ has a modular architecture.
It relies on a product of several numerical domains, which can be plug in and 
out. They exchange information via configurable reductions.
It also features a parameterisable abstract iterator tailored for flow- and
context-sensitive
analysis, and trace partitioning to achieve partial path-sensitivity.
In previous work \cite{magic2}, it has been specialised towards the
analysis of embedded avionics software by incorporating adapted iterations
strategies and numerical domains 
(such as relational octagons \cite{mine:oct} and domain-specific filter 
domains \cite{filters}).
However, its memory model was limited to simple well-structured data only, which
was sufficient at that time.
In order to analyze new code featuring union types and pointer casts,
we replaced it with our new memory abstractions.
Thanks to the modular construction of \astree\ and its modular proof of
correctness, most parts were not tied to the old memory
abstraction and could be reused (in particular, all numerical and partitioning
domains, as well as the iterator).

\subsection{Preliminary Experimental Results}

\def\mystrut{\hbox{\vrule height9pt depth0pt width0pt}}

\begin{figure}\begin{center}
\begin{tabular}{|r|r|r|r|r|c|}
\cline{2-6}
\multicolumn{1}{c}{\mystrut} & 
\multicolumn{2}{|c}{old domain} & 
\multicolumn{2}{|c|}{new domain} & 
\multicolumn{1}{|c|}{both} \\
\hline
\multicolumn{1}{|c}{lines\mystrut} & 
\multicolumn{1}{|c}{time} & 
\multicolumn{1}{|c}{mem.} & 
\multicolumn{1}{|c}{time} & 
\multicolumn{1}{|c}{mem.} & 
\multicolumn{1}{|c|}{alarms} \\
\hline
  9 500 &   82s & 0.2GB &   99s & 0.2GB & 1\mystrut\\
 70 000 &   62m & 1.0GB &   63m & 1.1GB & 0\\
226 000 &  4h57 & 1.6GB &  4h42 & 1.7GB & 1\\
400 000 & 11h04 & 3.0GB & 11h46 & 3.2GB & 0\\
\hline
\end{tabular}
\caption{Regression tests for \astree.}
\label{regresfig}
\end{center}\end{figure}

\begin{figure}\begin{center}
\begin{tabular}{|c|r|r|r|c|}
\hline
source & \multicolumn{1}{c|}{lines} & \multicolumn{1}{c|}{time} & 
\multicolumn{1}{c|}{memory} & alarms\mystrut\\
\hline
\multirow{2}{*}{end-user 1} 
& 35 000 & 12m & 212MB & 22\mystrut\\     
& 46 000 & 16m & 271MB & 84\\             
\hline
\multirow{2}{*}{end-user 2} 
&  92 000 & 3h17 & 3.2GB & 71\mystrut\\  
& 184 000 & 4h55 & 1.1GB & 36\\          
\hline
\end{tabular}
\caption{Four newly analyzed codes, from two end-users.}
\label{benchfig}
\end{center}\end{figure}

We have run three kinds of experiments: small case studies, regression tests 
and preliminary analyses of new real-life software.
They all ran on a 64-bit AMD Opteron 250 (2.4GHz) 
workstation, using one processor.
The analyzed programs do not feature recursion, dynamic memory 
allocation, nor multi-threading. Moreover, they are self-contained: they do 
not call precompiled library routines, and the external environment
is modeled using {\em volatile\/} variables.

Firstly, we tested the relevance of our domains to the specific problems
of union types and pointer casts discussed in Sect.~\ref{motivsect}.
We produced and were given by end-users several constructed
programs of a hundred lines, in the spirit of Figs.~\ref{msgex}--\ref{emuex}.
We were able to prove the absence of run-time errors of all case studies
in a fraction of second.

Secondly, we re-analyzed the pointer- and union-free industrial embedded 
critical code successfully analyzed by \astree\ in previous work \cite{magic2}.
Fig.~\ref{regresfig} compares the performance of the old and new memory domains.
We see that the memory peak and time consumption are only slightly increased, 
in the worse case, and we find the same alarms.
Note that, as \astree\ uses incomplete methods---such as partially
reduced products and convergence acceleration---there is no theoretical 
guarantee that our new memory semantics always gives more precise results than 
the former one, even though it is more expressive.
Hence, the importance of asserting experimentally non-regression in terms of
precision.

Thirdly, we analyzed four new industrial critical embedded software featuring 
unions and complex pointer manipulations.
Such codes could not be considered before in \astree\ because of its limited
legacy memory domain.
The analyses results are shown in Fig.~\ref{benchfig}.
These results are preliminary in the sense that we have not yet investigated
the causes of all alarms: they may be due to analyzer inaccuracies, but also
to real errors or too conservative assumptions on the environment.
The results are encouraging: they correspond to the preliminary results
obtained on the codes of Fig.~\ref{regresfig} before domain-specific
numerical domains and iteration strategies were incorporated in
\astree\ to achieve zero alarm \cite{magic2}.


\section{Related Work}
\label{relsect}

Several dialects of C, such as CCured \cite{ccured}, have been proposed to 
prevent error-prone uses of unions and pointers.
The value analysis of such dialects, with their cleaner memory
model, would be easier than the full C.
Unfortunately, their strengthened type systems would reject constructs
found legitimate by end-users and force them to rewrite their software.
For now, we (analysis designers) should adapt our analyses to the programming 
features they currently use.

There exists a very large body of work concerning pointer analyses for 
C---we refer the reader to the very good survey by Hind \cite{survey}.
Unfortunately, they cannot serve our purpose.
All field-insensitive methods natively support union and pointer 
casts---they are considered ``no-op.'' 
However, in order to find precise bounds on values stored into and then
fetched from memory, we absolutely require field sensitivity.
Very few field-sensitive analyses support unions or casts.
Most of them---{\em e.g.}, the recent work of Whaley and Lam \cite{pt}---assume
a memory model {\em \`a la\/} Java, where the memory can be 
{\em a priori\/} partitioned into cells of unchanging type.
As a middle-ground, Yong et al. \cite{horwitzbis} propose to collapse
fields upon detecting accesses through pointers whose type
mismatches the declared type of the fields.
This is not sufficient to treat precisely union types---Fig.~\ref{msgex}---or
polymorphism---Fig.~\ref{memcpyex}.
Also, flow-insensitive analyses (such as the union- and cast-aware
analysis by Steensgaard \cite{steensgaard}) which are 
well-suited for program optimization and understanding, would not perform
precise-enough for value analysis.
Indeed, they tend to produce large points-to sets---especially 
given that we are field-sensitive---which results in weak updates and 
precision losses in the numerical domains.
When it comes to program correctness, we are ready to use much more costly
abstractions: each instruction proved correct automatically 
saves the user an expensive manual proof.

Instead of relying on the structure of C types, 
we chose to represent the memory as flat sequences of bytes.
This allows shifting to a representation of pointers as pairs:
a symbolic base and a numeric offset. It is a common practice---it
is used, for instance, by Wilson and Lam in \cite{wilson}.
This also suggests combining the pointer and value analyses into a single 
one---offsets being treated as integer variables.
There is experimental proof \cite{pioli99} that this is more precise
than a pointer analysis followed by a value analysis.
Some authors rely on non-relational abstractions of offsets---{\em e.g.},
a reduced product of intervals and congruences \cite{repsx86}, 
or intervals together byte-size factors \cite{horwitz}.
Others, such as \cite{venet,rinard} or ourself, permit
more precise, relational offset abstractions.

We stress on the fact that using an offset-based pointer representation solves,
by itself, the problem of points-to analysis in the presence of
union types and casts, but it does not solve the problem of analyzing 
precisely the {\em contents\/} 
of the memory such offset-based pointers point to.
Several kinds of solution have been used to avoid treating this second problem.
A first one is to perform the field-sensitive points-to and value analysis 
of only a part of the memory that is never accessed through 
casts---{\em e.g.}, the {\em surface structure\/} of 
\cite{venet}---while the rest is only checked for in-bound accesses.
A second one is to fix one memory layout---using, {\em e.g.}, the
declared variable types or some pointer alignment constraints---and 
conservatively assume that mismatching dereferences result in any value 
\cite{repsx86}.
A less conservative solution, proposed by Wilson and Lam \cite{wilson}, 
is to consider that a dereference can output the value of any overlapping
cell.
We are more precise and more general because we allow value recomposition 
form individual bytes of partially overlapping cells and take into account the 
bit-representation of types.
In particular, unlike previous work, we can analyze precisely the indirect 
dereferencing following the memory copy of Fig.~\ref{memcpyex}.
Moreover, while \cite{wilson} often resolves a dereference into several 
overlapping cells, even when the target of the dereference is precisely known, 
we manage to select a single cell most of the time. 
This reduces the possibility of weak updates and improves 
the analysis precision,
especially when using relational numerical domains.
To our knowledge, our method is the first one that allows discovering 
precise {\em relational\/} invariants in the presence of union types and
pointer casts.

Finally, note that most articles---\cite{venet} being a notable
exception---directly leap from a memory model informally described in English 
to the formal description of a static analysis.
Following the Abstract Interpretation framework, we give a full
mathematical description of the memory model before presenting computable
abstractions proved correct with respect to the model.


\section{Future Work}
\label{futuresect}

A first goal is to reduce the number of alarms in the newly
analyzed codes of Fig.~\ref{benchfig}.
In the best scenario, most inaccuracies will be solved by tweaking already
existing parameters---such as the level of path sensitivity or domain 
relationality.
However, we will probably also need to add new numerical domains in the reduced
product $\D^\s_\R$, as it was necessary
in order to achieve the proofs of absence of run-time 
errors in \cite{magic2}.
We plan to investigate particularly the numerical domains required
to abstract pointer offsets precisely, as it is a new requirement of our
memory abstractions.
Finally, by iterating the analyzer refinement process over other codes
involving unions and pointers, we hope to provide a library of
abstractions that, in practice, is sufficient to analyze a large class of
embedded C programs.

Further goals include incorporating domains for heap-allocated
objects---{\em e.g.}, related to predicate-based summarization as proposed
by Sagiv et al. \cite{tvla,weak}.
We also wish to include other memory abstractions within our framework,
for instance, the string abstraction by Dor et al. \cite{dor} as well as
generalizations of $\D^{\s\mr{Eq}}_M$ using predicate 
abstractions parameterized by numerical domains
{\em \`a la\/} Cousot \cite{paramabs}.


\section{Conclusion}
\label{conclusionsect}

In this article, we proposed new techniques to perform the precise
value analysis of C programs with pointers and union types.
We first gave a precise meaning to such programs by defining a concrete memory 
semantics, parameterized by an Application Binary Interface.
We then proposed two computable abstractions: a value abstraction, 
parameterized by the choice of a numerical abstract domain, and an equality 
predicate abstraction, able to precisely deal with polymorphic memory copies.
The combined abstractions have been implemented within the 
\astree\ parametric static analyzer that checks for run-time errors in 
embedded critical C software.
Preliminary experimental results are encouraging: while not sacrificing the
precision and efficiency of \astree\ on legacy analyses---in particular,
the proof of absence of run-time errors for some large industrial codes
in a few hours of computation time---we greatly enlarge the class of
analyzable programs.
Currently, small test cases containing pointers and unions have been proved
correct while there are still a few dozens alarms on real-life industrial
examples.
We are confident that these results will be improved in the future
by refining the analyzer.


\appendix

\acks

We would like to thanks the whole \astree\ team
\cite{magic2}, as well as the anonymous
referees for their insightful comments.



\end{document}